\documentclass{aastex}           
\usepackage{graphicx}
\usepackage{natbib}
\begin{document}
\title{The size and shape of the oblong dwarf planet Haumea}

\author{Alexandra C. Lockwood\altaffilmark{1}, Michael E. Brown\altaffilmark{1} and John Stansberry\altaffilmark{2}}

\altaffiltext{1}{Division of Geological and Planetary Sciences, California Institute of Technology, Pasadena, CA 91125}
\altaffiltext{2}{JWST/NIRCam Team, Space Telescope Science Institute, Baltimore, MD 21218}
              \email{alock@caltech.edu}

\begin{abstract}
We use thermal radiometry and visible
photometry to constrain the size, shape, and albedo of the large
Kuiper belt object Haumea. The correlation between the visible and
thermal photometry demonstrates that Haumea's high amplitude and
quickly varying optical light curve is indeed due to Haumea's extreme
shape, rather than large scale albedo variations.  However, the well-sampled high precision visible data 
we present does require longitudinal surface heterogeneity to account for the shape of lightcurve.
The thermal emission from Haumea is consistent with the expected
Jacobi ellipsoid shape of a rapidly rotating body in hydrostatic equilibrium.
The best Jacobi ellipsoid fit to 
the visible photometry implies a triaxial ellipsoid
with axes of length 1920 x 1540 x 990 km
and density 2.6 g cm$^{-3}$, as found by \citet{Lellouch10}.  
While the thermal and visible data cannot uniquely constrain 
the full non-spherical shape of Haumea, 
the match between the predicted and measured thermal flux 
for a dense Jacobi ellipsoid suggests that Haumea is 
indeed one of the densest objects in the Kuiper belt.
\end{abstract}

\section{Introduction}
\label{sec:intro}
Haumea, one of the largest bodies in the Kuiper Belt, is also one of the
most intriguing objects in this
distant population.  Its rapid rotation rate, multiple satellites, and
dynamically-related family members all suggest an early giant impact
\citep{Brown07}.  Its surface spectrum reveals a nearly pure water
ice surface \citep{Barkume06,Trujillo07,Merlin07,Dumas11}, with constraints on other organic compounds 
with upper bounds $<$ 8\% \citep{Pinilla09}.  \citet{Lacerda08} also find evidence for a dark spot on one side of the rotating body, make the surface albedo non-uniform.  Even more interesting is that shape modeling has suggested a
density higher than nearly anything else known in the Kuiper belt and
consistent with a body almost thoroughly dominated by rock \citep{Rabinowitz06,Lacerda08,Lellouch10}.
\citet{Lacerda07} concluded a density of 2.551 g cm$^{-3}$ and follow up work found consistent values between 2.55 and 2.59 g cm$^{-3}$, depending on the model used.  Such a rocky body with an icy exterior could be a product of initial
differentiation before giant impact and subsequent removal of a significant
amount of the icy mantle.  \citet{Leinhardt10} demonstrate that such an impact is possible in a graze
 and merge collision between equal size bodies, and are able to reproduce the properties of the Haumea family system.

Much of our attempt at understanding the history of Haumea relies on the
estimate of the high density of the body.  Haumea's large-amplitude light curve and
rapid rotation have been used to infer an elongated shape for the body.  
Assuming that the rotation axis lies in the same plane as the plane of
the satellites, the amplitude of the light curve then gives a ratio of
the surface areas along the major and minor axes of the bodies. If the
further assumption is made that Haumea is large enough to be in
hydrostatic equilibrium, the full shape can be uniquely inferred to be a
Jacobi ellipsoid with fixed ratios of the three axes.  From this shape
and from the known rotation velocity, the density is precisely
determined.  Finally, with the mass of Haumea known from the dynamics of
the two satellites \citep{Ragozzine09}, the full size and shape of Haumea
is known \citep{Rabinowitz06,Lacerda08,Lellouch10}.

The major assumption in this chain of reasoning is that Haumea is a
figure of equilibrium.  While it is true that a strengthless
body can instantaneously have shapes very different from figures of
equilibrium \citep{Holsapple07}, long-term deformation at the pressures
obtained in a body this size should lead to essentially fluid behavior,
at least at depth.
Indeed, any non-spinning body large enough to become round due to
self-gravity has attained the appropriate figure of equilibrium. While
the size at which this rounding occurs in the outer solar system is not
well known, the asteroid Ceres, with a diameter of 900 km is essentially
round \citep{Millis87,Thomas05}, while among the icy
satellites, everything the size of Mimas and larger ($\sim$400 km)
is essentially round.
It thus seems reasonable to assume that a non-spinning Haumea, with a
diameter of $\sim$1240 km \citep{Lellouch13,Fornasier13} would be round, thus a rapidly-spinning Haumea
should likewise have a shape close to that of
the Jacobi ellipsoid defined by Haumea's density and spin rate.

Nonetheless, given the importance of understanding the interior structure of Haumea and the unusually high density inferred from
these assumptions, we find it important to attempt independent size and
density measurements of this object.
Here we use unpublished photometric data from the Hubble Space Telescope (HST)
to determine a best-fit Jacobi ellipsoid model. We then compare the
predicted thermal flux from this best-fit Jacobi ellipsoid to 
thermal flux measured from the Spitzer Space Telescope(first presented in \citet{Stansberry08}) and
consider these constraints on the size and shape of the body.

\section{Observations}
\label{sec:obs}
\begin{figure}
\includegraphics[scale=0.6]{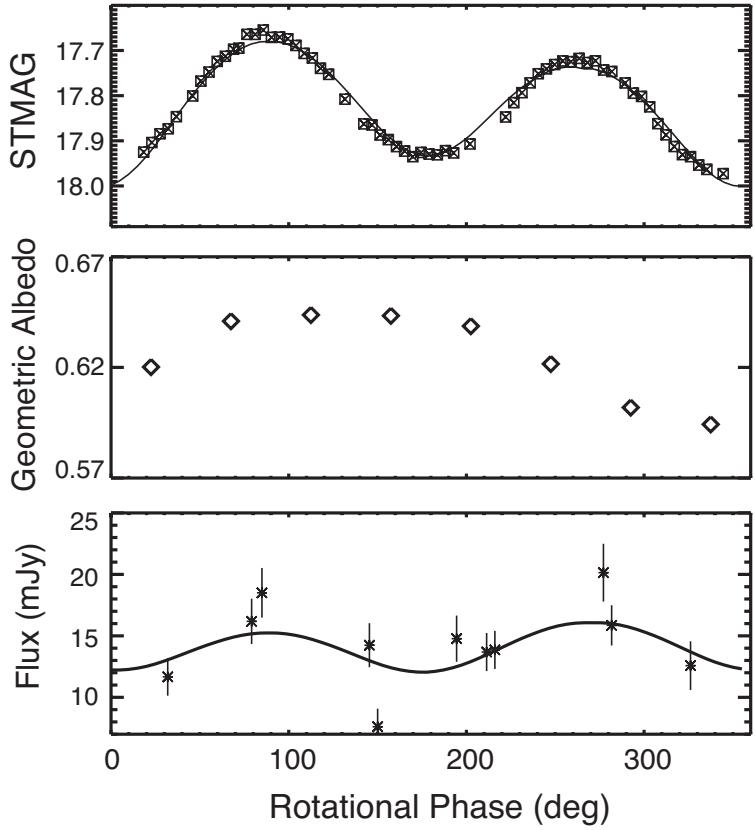}
\caption{The visible lightcurve, geometric albedo, and thermal lightcurve plotted over one rotation.  The error bars in the top two panels are smaller than the size of the plotted point.  The visible photometry are
fit with a Jacobi ellipsoid of dimensions 
1920 x 1540 x 990 km with the modest longitudinal variation in
reflectance shown. This ellipsoid model provides a quite good fit 
to the 70 $\mu$m thermal data from Spitzer.}
\label{fig:lightcurves}
\end{figure}

Haumea was imaged on 2009 February 4 using the PC chip on the Wide
Field/Planetary Camera 2 on HST.  We obtained 68 100s exposures using the
F606W filter, summarized in Table 1. The observations were obtained over 5 consecutive HST orbits, which 
provides a full sample of Haumea's 3.9154 hr rotational lightcurve \citep{Rabinowitz06}.
With a pixel scale of $\sim$5500 km at the distance of Haumea and semi-major axes of $\sim$50,000 km and $\sim$25,000 km for Hi'iaka and Namaka, respectively \citep{Ragozzine09}, this is the first published dataset where the object is resolved from
its satellites, providing a pure lightcurve of the primary.
During the observation both satellites are sufficiently spatially
separated from the primary that we are able to perform circular aperture
photometry.  Basic photometric calibrations are performed on the data
including flat fielding, biasing, removing charge transfer efficiency
effects, and identifying and removing hot pixels and cosmic
rays\footnote{see the WFPC2 handbook at http://www.stsci.edu/instruments/wfpc2/Wfpc2\_dhb/ WFPC2\_longdhbcover.html}. 
In five of our images the primary is
contaminated by cosmic rays so we do not include these data.  
A 0.5" aperture is used to measure the object, and we apply an infinite
aperture correction of 0.1 magnitudes  \citep{Holtzman95}.  We
present and model the data in the STMAG magnitude system, but a
convolution of the F606W filter with the Johnson V filter shows a
difference of approximately 0.1 magnitudes. This difference, along with
the satellite flux contributions of ~10\% \citep{Ragozzine09} included in previous photometry of the dwarf planet,
indicate that the magnitude and amplitude of the lightcurve presented here($\Delta$m = 0.32) is
consistent with the findings of \citet{Rabinowitz06} ($\Delta$m = 0.28) and \citet{Lacerda08} ($\Delta$m = 0.29). 
The rotational period of Haumea is known sufficiently precisely that all of the 
observations are easily phased. We combine our data with that of  
\citet{Rabinowitz06} and \citet{Lacerda08} to get a 4-year baseline of observations and find a period of $3.91531 \pm 0.00005$ hours using phase dispersion minimization \citep{Stellingwerf78}.
This period is consistent with that of \citet{Lellouch10} (P = 3.915341 $\pm$ 0.000005 h derived from a longer baseline), whose more precise solution we use to phase all observations, visible and infrared.  With this period, there is a 60s uncertainty in the phasing over the 1.5 years between observations.  We define a phase of 0 to be the point of absolute minimum brightness of Haumea 
and define a longitude system in which $\lambda = [(JD-2454867.042]$ modulo 360 degrees.  This result is given in JD at Haumea, which is consistent with the phased data of \citet{Lacerda08}, and \citet{Lellouch10} who instead quote their phasing in JD at the Earth. The photometric results are shown in Fig. \ref{fig:lightcurves}. 

The thermal radiometry of Haumea was obtained 2007 July 13 -- 19 using
the 70 micron band of the MIPS instrument \citep{Rieke04} aboard the
Spitzer Space Telescope (SST). Haumea's lightcurve was unknown at the time, and the objective of the observations was simply to detect the thermal emission at a reasonably high signal to noise ratio (SNR). The data were collected as three 176 minute long observations, each of which is nearly as long as the (now known) lightcurve period. 
\citet{Stansberry08} published the flux obtained by combining all three
observations, as well as models indicating a diameter of about 1150 $\pm$ 175 km.
Contemporaneously, Haumea's lightcurve was published \citep{Lacerda08},
a result that led us to consider re-analysing the Spitzer data to try and detect
a thermal lightcurve. To that end, we split the original 176 minute exposures
(each made up of many much shorter exposures) into 4 sub-observations, each
44 minutes long. The resulting data was processed using
the MIPS Instrument Team pipeline \citep{Gordon05}, resulting in
flux calibrated mosaics for each of the 12 sub-observations.  These individual observations are presented in Table 2.  One of the points was obviously discrepant and was removed from the analysis.
The data were also re-processed using improved (relative to
the 2007 processing published in \citet{Stansberry08}) knowledge of Haumea's 
ephemeris. The reprocessing was undertaken as part of a project to reprocess all 
Spitzer/MIPS  observations of TNOs (described in \citet{Mommert12})
and is key to obtaining the highest SNR from the data. Had Haumea's lightcurve
been known at the time the Spitzer observations were planned, the observations
probably would have been taken, for example, as a series of about 10 approximately 
60 min exposures spaced about 4.1 rotations apart. The non-optimal observation
plan may in part explain why the the uncertainties on the flux measurements
appear to be somewhat optimistic, as discussed in more detail below.

The motion of Haumea was significant over the 6-day observing interval,
so we were able to make a clean image of the background sources (i.e. one
without contamination from Haumea), and then subtract that sky image
from our mosaics. The procedure used has been described previously, e.g.
by \citet{Stansberry08}. We performed photometry on the sky-subtracted
images, obtaining significantly smaller uncertainties than was possible using
the original mosaics. The raw photometry was corrected for the size of
the photometric aperture (15'' radius).
The signal-to-noise ratio of the resulting detections was about 7 in each of the
12 epochs. An additional calibration uncertainty of 6\% should be systematically
applied to the entire dataset. The thermal results are shown in Fig. 1.
Where multiple observations are made at the same phase, these observations
are averaged, and the uncertainty is taken from the standard deviation
of the mean (or the full range if only two points go into the mean).

\begin{figure}
\includegraphics[scale=0.6]{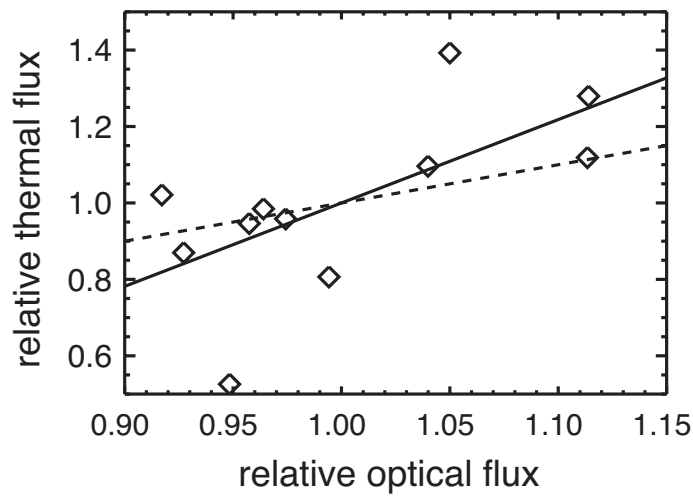}
\caption{The correlation of the relative optical and thermal flux. The dashed line
shows a one-to-one correlation while the solid line shows the best fit. A
rank correlation test shows that the two distributions are correlated
at the 97\% confidence level. The in-phase thermal light curve of
Haumea demonstrates that it is an elongated body.}
\label{fig:correlation}
\end{figure}

Though the uncertainties are large, the thermal light curve 
appears in-phase with the measured visible light curve. 
To robustly ascertain the detection of a thermal lightcurve, 
we look for a correlation between the thermal and visible datasets. 
We compare the mean visible flux
during the phase of each 44 minute long thermal observation with the
measured thermal flux during that observation (Fig. \ref{fig:correlation}). 
A linear fit to the data suggests that a positive correlation between the optical and thermal brightness.  
As noted above, the deviation of the measurements from the model are larger than expected, so we assess the significance of the correlation between the optical and thermal data using
the non-parametric Spearman rank correlation test \cite{Spearman1904}. With
this test, we
find that the two data sets are correlated at the 97\% confidence level,
and that the correlation is positive, that is: the observed optical
and thermal flux
increase and decrease in phase and there is only a 3\% chance that this phase correlation
is random. 
This positive correlation between the thermal and visible data sets indicates that
we are viewing an elongated body and that the visible light curve must be caused -- at 
least in part -- by the geometric effects of this elongated body. 

\section{Photometric model}
\label{sec:phot}
We begin with the assumption that Haumea is indeed a Jacobi ellipsoid whose shape
is defined by its density and spin period.
To find the Jacobi ellipsoid which best fits the photometric data,
we model the expected surface reflection from an ellipsoid
by creating a mesh of 4,000 triangular facets
covering the triaxial ellipsoid and then determining the
the total visible light from the sum of the light reflected back 
toward the observer from each facet. Facets are approximately equal-sized equilateral triangles
with length equal to 5 degrees of longitude along the largest circumference of the body.
Mesh sizes a factor of 2 larger or smaller give identical results.
For each facet we use a Hapke photometric model \citep{Hapke93} to determine the 
reflectance as a function of emission angle. This model accounts for the effects of low phase angle 
observations, such as coherent backscattering and shadow hiding, and has been used to model the
 reflectance of many icy surfaces. 
For concreteness, we adopt parameters determined by \citet{Karkoschka01} for
Ariel, a large satellite which exhibits deep water ice
absorption and a high geometric albedo. We utilize the published values for the mean surface
 roughness($\bar{\theta} = 23^\circ$), single-scattering albedo ($\varpi$ = 0.64), asymmetry parameter($g = -.28$), 
and magnitude($S(0)$) and width($h$) of the coherent backscattering and shadowing functions, 
$S(0)_{CB} = 4.0$, $h_{CB} = 0.001$, $S(0)_{SH} = 1.0$, and $h_{SH} =0.025$ \citep{Karkoschka01} . 
While we have chosen Ariel because
it is perhaps a good photometric analog to Haumea,
we do note that within the range of Hapke parameters of icy objects throughout the
solar system ($\varpi \sim 0.4 \-- 0.9$, $g = -.43 \-- -.17$, $\bar{\theta} = 10 \-- 36^\circ$), including 
the icy Galilean satellites and Triton \citep{Buratti95,Hillier90}, the precise parameters chosen will affect only the geometric albedo and the beaming parameter, as discussed below.  

The visible flux reflected from the body then becomes:

\begin{equation}
F_{vis} = p_{vis} \frac{F_{\odot, 606}}{{R_{AU}}^2} A \frac{H \cos e}{\pi \Delta^2} \nonumber 
\end{equation}

where $p_{vis}$ is the visual albedo, $ F_{\odot,606}$ is the solar luminosity over the bandwidth of the F606W filter, $A$ is the projected surface area, $H$ is the Hapke reflectance function, $e$ is the angle of incidence, $\Delta$ is the geocentric distance to the body, and $R$ is the heliocentric distance.

Our modeled body is rotated about the pole
perpendicular to our line of sight -- consistent with the hypothesis 
that the rotation pole is similar to the orbital pole of the satellites -- and the photometric light
curve is predicted. For such a model, the peak and trough of the visible
lightcurve correspond to the largest and smallest cross-sectional areas
of the body, and the ratio of the length of the largest non-rotational axis ($a$)
to the smallest non-rotation axis ($b$) controls the magnitude of the photometric
variation.  With a uniform albedo across the
surface of Haumea, however, no triaxial ellipsoid can fit the asymmetric observed
lightcurve.  We confirm the assertion of \citet{Lellouch10}
that the photometric variations of Haumea are caused primarily by shape and 
that surface albedo variations add a only minor modulation.
In this approximation, the brightest peak and brightest trough of the data
are assumed to be from essentially uniform albedo surfaces and are modeled
to determine the ratio of the axes of the body. For plausible values
of the dimension of the rotational axis ($c$), the measured peak-to-trough amplitude in the lightcurve 
of $\Delta m = 0.32 \pm 0.006$ is best modeled with an axis ratio of $b/a =0.80 \pm 0.01$. Using the brightest trough and darkest peak instead, the axial ratio would be b/a = 0.83, but the surface heterogeneity necessary for this assertion is less likely than the single darker spot proposed here and observed by others \citep{Lacerda08}. %Due to the precision of the 
%visible dataset and the realistic surface conditions we model, we can assert that this is indeed the true axial ratio.

Our measured value of 0.80 differs from previous measurements of 
$b/a$ of .78 \citet{Rabinowitz06} and .87 \citet{Lacerda08} for a number of reasons.  Our 
visible dataset resolves Haumea from its satellites 
which results in a slightly deeper lightcurve.  More to the point, including a
realistic surface reflectance model changes that estimated shape significantly.  \citet{Rabinowitz06}
 do not actually model the shape of the body, while \citet{Lacerda08} assume the surface 
 to be uniformly smooth, giving a value for $b/a$ that is too high.
 More recently, \citet{Lellouch10} confirm a more elongated body ($b/a$ = .80),
 after having tested two different models, including that of \citet{Lacerda08}. 

With the ratio of the axes fixed, we now
find the simplest surface normal albedo model consistent with the data. We
divide the surface longitudinally into 8 slices and allow the albedo to vary
independently between sections to account for the possible hemispherical
variation apparent in our data and explored by others \citep{Lacerda08,Lacerda09}.  
Dividing the surface further does not significantly improve the fit.
The middle panel of Figure \ref{fig:lightcurves} shows how the geometric 
albedo varies across the surface of the body.

Assuming that Haumea is indeed a Jacobi ellipsoid, the ratio $b/a=0.80$ 
combined with the rotation period uniquely defines $c/a=0.517$ and a density of
2.6 g cm$^{-3}$. Combining these parameters with the known mass of Haumea from \citet{Ragozzine09} implies
$a= 960$ km, $b= 770$ km, and $c=495$ km. These radii agree with the ones obtained by \citet{Lellouch10} using a Lommel-Seelinger reflectance function.

\section{Thermal model}
\label{sec:therm}
To calculate the thermal emission from our shape and albedo
model we first determine the temperature of each facet of the body.
Due to Haumea's rapid rotation, the temperature of any given face of
the surface does not have time to equilibrate with the instantaneous
incoming insolation.  Instead we calculate the average
amount of sunlight received by a facet during a full rotation, which is only dependent
on the angle between rotational pole and the facet normal.
Although we implement a standard thermal model, the rapidly spinning object gives rise to surface temperatures indicative of an isothermal latitude model \citep{Stansberry08}, which we precisely calculate for the non-spherical geometry of the object.  Indeed the agreement between the visible and infrared datasets as seen in Figure \ref{fig:correlation} supports the hypothesis of a body with negligible thermal conductivity.

We model the average amount of
sunlight absorbed for each facet by multiplying the geometric albedo of each
facet by an effective phase integral $q$ and averaging over a full
rotation. The average facet phase integral is a mild function
of the shape of the body, but for simplicity we simply adopt a value of $q=0.8$, 
as used by \citet{Stansberry08} for large, bright KBOs. In fact, the precise value
used has little impact on our final results.

If we assume the surface is in thermal equilibrium, the temperature of each facet, is determined by balancing this absorbed sunlight
with the emitted thermal radiation. We choose a typical thermal
emissivity of 0.9 and invoke a beaming parameter, $\eta$, which is a simple correction to the total amount of
energy radiated in the sunward direction, usually assumed to be caused
by surface roughness, but which can be taken as a generic correction factor
to the assumed temperature distribution. For asteroids of known sizes, 
\citet{Lebofsky86} found $\eta$ to be approximately 0.756, a correction
which agrees well with measurements of icy satellites in the outer solar
system \citep{Brown82a,Brown82b}.  The beaming parameter value range for 
Trans-Neptunian objects is fully described in \citet{Lellouch10}, who found $\eta$ of 1.15-1.35 
for Haumea, with hemispherical variations consistent with a much lower value ($\eta \sim$ .4-.5). 
The MIPS and PACS fluxes values presented in \citet{Lellouch10} were updated and presented in 
 \citet{Fornasier13}, who, also using the SPIRE data, found a beaming factor of 0.95$^{+0.33}_{-0.26}$ with a NEATM model.
 We leave this as a free parameter in our modeling but consider an inclusive range.

For each rotational angle of our model, we predict the total thermal
flux by calculating the blackbody spectrum from each visible facet and 
integrating this flux in the full 70 $\mu$m band pass of MIPS, according to

\begin{equation}
\label{eq:IR}
F_{IR} = \frac{A \cos e}{\pi^2 \Delta^2}\epsilon \int{B_{\lambda} (T(\theta, \phi)) \sin \theta d\theta d\phi} \nonumber
\end{equation}

where $A$ is the cross-sectional surface area, $\epsilon$ = 0.9 is the emissivity, $B$ is the Planck function and $T$ is the temperature at each piece of the surface.  Assuming a solar flux at 70 $\mu$m of $S$ at the distance of Haumea, $R$, then $T$ is calculated for an edge-on rotating body and 

\begin{equation}
\label{eq:temp}
 T = [\frac{S*(1-qp_{vis})}{\epsilon \sigma \eta R^2}]^{1/4} \nonumber 
\end{equation}

Figure \ref{fig:lightcurves} shows the measured Spitzer flux along with the flux predicted
from a model for the theoretical Jacobi ellipsoid with $a=960$ km and $a:b:c = 1.00:0.80:0.52$ 
and a thermal beaming parameter of $\eta=0.76$. 
The best fit is obtained by assuming $\eta=0.89$, but values of $\eta$ between 0.82 and 0.97
are within the 1-$\sigma$ error limits.  The larger values measured by \citet{Lellouch10} and quoted for the majority of Kuiper Belt objects ($\sim$1.2 by \citet{Stansberry08} and $\sim$ 1-2.5 by \citet{Lellouch13}) are consistent with our result if we consider the difference of thermal models employed.  \citet{Stansberry08} explain that the surface temperature difference between an isothermal latitude model (that is used in this work) and a standard thermal model (used by \citet{Lacerda09} and \citet{Lellouch10} is simply a factor of $\pi^{-\frac{1}{4}}$.  If this value is incorporated into the beaming factor, the inconsistency between the models and resulting beaming parameters is resolved.
Remarkably, the shape and albedo model constructed from only
photometric observations  and the assumption of fluid equilibrium
provides an acceptable prediction of the total thermal flux at 70 $\mu$m
and its rotational variation.

\section{Discussion}
\label{sec:disc}
The thermal and photometric light curves of Haumea are consistent with the
assumption that Haumea is a fluidly relaxed, rapidly rotating Jacobi ellipsoid
with a density of 2.6 g cm$^{-3}$ and minor albedo variation across its surface. 
Although we allow the albedo to change longitudinally, 
we have demonstrated that the reason 
for the double-peaked lightcurve of Haumea is in fact a shape effect. This work presents new data and an informed Hapke model that agree with the findings of previous authors \citep{Lellouch10,Lacerda08,Rabinowitz06}.  

The lightcurve in several colors \citep{Lacerda08,Lacerda09} indicates 
that the albedo variation is concentrated in a large spot on one side of the body, an argument
 supported by our allowed albedo variation.  The precise geometric albedo presented here is less than that of \citet{Fornasier13}, but with a small variation in either the single-scattering albedo or the asymmetry parameter, we easily find agreement between the two values.  We do not focus on the absolute value of the albedo here, as it is closely tied to the unknown Hapke parameters for the surface.  The only affect this has on the thermal fit is to change the best fit beaming parameter, which does not differ by more than 1-$\sigma$ from the reported value.  The important point here, however, is that our precise visible data set agrees with the dark spot proposed by \citet{Lacerda08}.
 
 There are still a number of questions remaining regarding this KBO.  The temperature of the dark spot is still uncertain and our data
 are not precise enough to constrain it, although the lower albedo we use to match the visible data agree well with a warmer region.  Another assumption used here that could be disputed is the rotation axis of the body as perpendicular to our line of sight.  However, assuming that the majority of the depth of the light curve is from the major axes of the body, the pole position will only affect the size of the third dimension of the body and the thermal beaming factor, which become somewhat degenerate when fitting the thermal data anyway.  If the body were not mostly edge-on, it would be difficult to explain the particular surface patterning needed to recreate the visible lightcurve. 
 
 While it is encouraging that a dense Jacobi ellipsoid fits both data sets,
it is important to point out that we can only obtain a unique solution 
under the assumption that the object has a prescribed shape. If 
Haumea is modeled as an arbitrary triaxial ellipsoid rather than a Jacobi
ellipsoid, large families of solutions are possible. To first order, the
photometric data constrain the ratio $b/a$, while the thermal flux is roughly
proportional to the emitting surface area which is proportional to $ac$ and $bc$.
As long as the ratio of $b/a$ is kept constant, however, equally good fits
can be obtained from very elongated ellipsoids with a very short rotation axis, or from
only moderately elongated ellipsoids with a large rotation axis, as long as
the value $ac$ is approximately constant. For this unconstrained problem, 
densities anywhere between 1 g cm$^{-3}$ (for very elongated objects)
and 3 g cm$^{-3}$ for compact objects are compatible with both data sets. 
While such large deviations from an equilibrium shape appear implausible, the thermal
data alone cannot rule them out.

\begin{table}
	\caption{HST F606W flux of Haumea}
	\label{tab:1}
	\begin{tabular}{cccccccc}
	\hline
	\hline
JD  & phase  &F606W  & uncertainty&
JD  & phase  &F606W  & uncertainty\\
-2450000 & (deg) & mag & & -2450000 & (deg) & mag &  \\
	\hline
4866.943802	&	142.27	&	17.863	&	0.004	&	4867.0898	&	104.08	&	17.689	&	0.003\\
4866.945802	&	146.87	&	17.865	&	0.004	&	4867.0918	&	108.68	&	17.707	&	0.003\\
4866.948202	&	151.47	&	17.887	&	0.004	&	4867.0937	&	113.28	&	17.717	&	0.003\\
4866.950202	&	156.06	&	17.897	&	0.004	&	4867.0962	&	117.88	&	17.740	&	0.003\\
4866.952102	&	160.66	&	17.913	&	0.004	&	4867.0981	&	122.47	&	17.753	&	0.003\\
4866.954602	&	165.26	&	17.923	&	0.004	&	4867.1025	&	131.67	&	17.808	&	0.004\\
4866.956502	&	169.86	&	17.935	&	0.004	&	4867.1435	&	222.08	&	17.847	&	0.004\\
4866.958502	&	174.45	&	17.926	&	0.004	&	4867.1455	&	226.68	&	17.815	&	0.004\\
4866.960402	&	179.05	&	17.930	&	0.004	&	4867.1474	&	231.28	&	17.793	&	0.004\\
4866.962902	&	183.65	&	17.931	&	0.004	&	4867.1494	&	235.87	&	17.772	&	0.004\\
4866.964802	&	188.24	&	17.922	&	0.004	&	4867.1518	&	240.47	&	17.751	&	0.003\\
4866.966802	&	192.84	&	17.927	&	0.004	&	4867.1538	&	245.07	&	17.741	&	0.003\\
4866.971202	&	202.04	&	17.907	&	0.004	&	4867.1557	&	249.66	&	17.731	&	0.003\\
4867.010702	&	289.39	&	17.772	&	0.004	&	4867.1577	&	254.26	&	17.723	&	0.003\\
4867.012702	&	293.98	&	17.795	&	0.004	&	4867.1601	&	258.86	&	17.725	&	0.003\\
4867.014602	&	298.58	&	17.802	&	0.004	&	4867.1621	&	263.46	&	17.718	&	0.003\\
4867.017102	&	303.18	&	17.825	&	0.004	&	4867.1640	&	268.05	&	17.727	&	0.003\\
4867.019002	&	307.77	&	17.862	&	0.004	&	4867.1660	&	272.65	&	17.723	&	0.003\\
4867.021002	&	312.37	&	17.887	&	0.004	&	4867.1684	&	277.25	&	17.743	&	0.003\\
4867.022902	&	316.97	&	17.913	&	0.004	&	4867.1704	&	281.85	&	17.747	&	0.003\\
4867.025402	&	321.57	&	17.931	&	0.004	&	4867.2143	&	18.39	&	17.925	&	0.004\\
4867.027302	&	326.16	&	17.935	&	0.004	&	4867.2163	&	22.99	&	17.904	&	0.004\\
4867.029302	&	330.76	&	17.954	&	0.004	&	4867.2182	&	27.58	&	17.884	&	0.004\\
4867.031202	&	335.36	&	17.963	&	0.004	&	4867.2202	&	32.18	&	17.873	&	0.004\\
4867.035602	&	344.55	&	17.973	&	0.004	&	4867.2226	&	36.78	&	17.847	&	0.004\\
4867.075202	&	71.90	&	17.695	&	0.003	&	4867.2265	&	45.97	&	17.801	&	0.004\\
4867.077102	&	76.50	&	17.664	&	0.003	&	4867.2285	&	50.57	&	17.768	&	0.004\\
4867.079602	&	81.10	&	17.664	&	0.003	&	4867.2309	&	55.17	&	17.748	&	0.003\\
4867.081502	&	85.69	&	17.655	&	0.003	&	4867.2329	&	59.77	&	17.724	&	0.003\\
4867.083502	&	90.29	&	17.671	&	0.003	&	4867.2348	&	64.36	&	17.713	&	0.003\\
4867.085402	&	94.89	&	17.671	&	0.003	&	4867.2368	&	68.96	&	17.697	&	0.003\\
4867.087902	&	99.49	&	17.674	&	0.003	&							
\end{tabular}
\end{table}
	
\begin{table}
	\caption{Thermal Flux of Haumea}
	\label{tab:2}
	\begin{tabular}{cccc}
	\hline
	\hline
JD - 2450000 & phase (deg) & MIPS 70 $\mu$m flux (mJy) & uncertainty (mJy)\\
	\hline	
4294.6770	&194.5&	14.77	&	1.88\\
%4294.7068	&260.2&	11.77	&	1.42\\ %****
4294.7367	&326.1&	12.58	&	1.98\\
4294.7664	&31.9&	11.67	&	1.53\\
4297.4004	&84.9&	18.51	&	2.02\\
4297.4302	&150.0&	7.61		&	1.47\\
4297.4601	&216.0&	13.86	&	1.55\\
4297.4899	&281.8&	15.86	&	1.64\\
4300.4980	&79.1&	16.18	&	1.84\\
4300.5277	&145.4&	14.24	&	1.79\\
4300.5576	&211.4&	13.69	&	1.55\\
4300.5874	&277.1&	20.14	&	2.35\\
	\end{tabular}
	\end{table}

\end{document}